# Hierarchical Design of Ambidirectional LCE Sheets for Three-State Gaussian Shape Morphs

Atalaya Milan Wilborn[1,2]; Gurminder K. Paink[1]; Raphael Kay[1]; Patryk Wąsik[3]; Friedrich Stricker[1]; Jacopo Movilli[1]; Shucong Li[4,5]; Joanna Aizenberg[1,5]*

1. John A. Paulson School of Engineering and Applied Sciences, Harvard University, Cambridge, MA 02138, USA.
2. Complex Materials, Department of Materials, ETH Zürich, Zürich, 8093 Switzerland.
3. National Synchrotron Light Source II, Brookhaven National Laboratory, Upton, New York 11973, USA.
4. School of Materials Science and Engineering, Georgia Institute of Technology, Atlanta, GA, USA.
5. Department of Chemistry and Chemical Biology, Harvard University, Cambridge, MA, USA.

*Corresponding author. Email: jaiz@seas.harvard.edu.

**Abstract**

Modulation of surface curvature is a commonly exploited strategy to achieve adaptability within natural and synthetic systems where structure governs function. While the Gaussian curvature of inextensible surfaces is fixed, surfaces made of extensible materials can adapt their Gaussian curvature under strain, unlocking capabilities such as locomotion, dynamic coloration, and tunable stiffness. This feature is commonly exploited to develop soft actuators capable of complex functionalities; however, traversing the entire spectrum of Gaussian curvature was previously difficult to achieve without the use of multi-material constructs or spatially varied stimulus fields. In this work, we present a strategy for generating a diverse range of dynamic Gaussian shape morphs using multiphase liquid crystal elastomers (LCEs) as ambidirectionally deforming building blocks. Taking advantage of the material's sequential contraction and elongation during phase transitions, we design and investigate the deformation response of radially aligned flat LCE sheets which consecutively morph into negatively and positively curved surfaces in response to monotonically increasing temperature. We then investigate the influence of mesoscale asymmetry and macroscale geometry on curvature magnitude and bending symmetry, thereby demonstrating that a single form factor can be programmed to undergo a rich library of deformation behaviors. Finally, inspired by the strategy of cephalopods to couple mechanical deformability to dynamic optical properties, we fabricate a structurally colored LCE sheet which couples its dynamic Gaussian curvature to a dynamic color-changing response. This approach offers a route toward adaptive structures capable of multistate shape reconfiguration for applications in soft robotics, deployable systems, and responsive surfaces.

## 1. Introduction

The morphing of 2D sheets into curved surfaces is ubiquitous in both natural and synthetic systems, from the growth-driven curling of leaves and petals [1–3] to the exploitation of bistable curved surfaces for deployable structures.[4,5] Indeed, surface curvature often governs a number of important mechanical, chemical, and biological properties within a system. For example, curvature alteration of extracellular matrices during tissue development shapes the migration patterns and function of cells growing within them[6,7]. Therefore, the ability to alter surface curvature is a highly utilized strategy for natural and synthetic systems to modulate properties and adapt to environmental changes. In synthetic soft systems, switchable surface curvature has been exploited to achieve tasks from extreme terrain exploration[8] to gripping.[9] The functionality of these actuators is determined by the compliance of their constituent materials, giving them a unique trait – extensibility. Inextensible materials preserve their Gaussian curvature ($K$) under bending, a rule intuitively exploited when holding a slice of pizza. By bending the initially flat slice along its crust, the pizza preserves $K$ by maintaining a straight line along its length, effectively preventing the pizza from drooping. However, soft materials are not subject to this constraint and can undergo significant changes to their curvature by stretching. Liquid crystal elastomers (LCEs) are an especially promising material for developing actuators with switchable $K$ because their stretches can be prescribed at the molecular level to generate complex shape morphs upon deformation.[10,11]

To generate out-of-plane deformations in nematic LCE sheets, previous studies have patterned mesogens into inhomogeneous in-plane director fields.[12–14] When these mesogens are radially aligned, contraction due to heating above the nematic to isotropic transition temperature ($T_N$) results in the sheet morphing into a hyperbolic paraboloid or saddle, where $K$ is negative everywhere.[12] Rotating this director field by 90° produces its orthogonal dual.[14] Now, contraction due to heating leads to the formation of a cone where $K$ is 0 everywhere except at the apex, where it is positive. Variations in radial distortion can lead to delocalization of Gaussian curvature, where $K$ is now positive everywhere and the sheet morphs into a dome[15]. Experimentally, the realization of director fields and their orthogonal dual pairs typically requires switching between heating and cooling LCE sheets to induce contraction and expansion, respectively.[12,14] However, actuation which relies on applying a stimulus above and below a reference state (i.e., heating and cooling beyond room temperature) offers practical challenges in terms of integration and stimulus control. Therefore, applications for dynamically curved actuators have been limited to switching using monotonically changing stimuli, typically heating[16–18] or irradiation.[9,19,20]

In this work, we demonstrate a diverse palette of oppositely curved surfaces obtained from magnetically aligned circular LCE sheets capable of ambidirectional deformations. Rather than relying on nonmonotonic stimulus exposure, these sheets use ambidirectional material building blocks to access multistate dynamic out-of-plane deformations under increasing temperature (*Fig. 1A, Fig. 1B*). The nature and magnitude of these deformed surfaces can be tuned as a function of both molecular alignment and macroscopic geometry. Finally, we use this ambidirectionality to enable three-state soft photonic actuation via structural coloration of a radially aligned LCE sheet. Together, these results advance the development of reconfigurable material platforms capable of coupling programmable shape change with functional responses across soft robotic, optical, and photonic systems.

## 2. Results and Discussion

### 2.1 Phase transition-driven ambidirectional Gaussian shape morphing

Recently, a family of LCEs were reported which pack into multiple LC phases, namely chevron Smectic C (cSmC) and Smectic A (SmA).[21] Macroscopically, inducing the cSmC-SmA-Iso phase transition results in sequential contraction and elongation along the axis of the director ($\hat{\mathbf{n}}$) (*Fig. 1B*). Taking advantage of this ambidirectional response, we aligned and polymerized LCE sheets comprised of LC monomer 4-(6-acryloxy-hex-1-yl-oxy)phenyl 4-(hexyloxy)benzoate, crosslinker 1,6-hexanediol diacrylate (HDDA) (2.5% wt/wt based on LC monomer) and photoinitiator 2-hydroxy-2-methylpropiophenone (2% wt/wt based on LC monomer) atop disk magnets to encode spatially varying radial director fields (*Fig. 1C, Fig. S1*). Upon heating, the sheet first undergoes radial contraction during the cSmC-SmA transition and corresponding circumferential expansion due to thermal Poisson effects.[14,22,23] This strain field generates negative $K$ everywhere, producing a saddle. With further heating, the strain field inverts due to the SmA-Iso phase transition. The resulting radial expansion and circumferential contraction lead to the formation of a dome with positive *K* everywhere. This deformation behavior was first predicted via finite element (FE) simulations (*Fig. 1D*) and subsequently confirmed experimentally (*Fig. 1E*).[21] Shape morphing across the entire spectrum of Gaussian curvature was previously hard to achieve in single-material structures without requiring nonmonotonic stimuli[11,13,24]. Ambidirectional LCE building blocks enable a single material to morph across the full spectrum of Gaussian curvature through a nonmonotonic strain response driven by monotonically increasing stimuli.

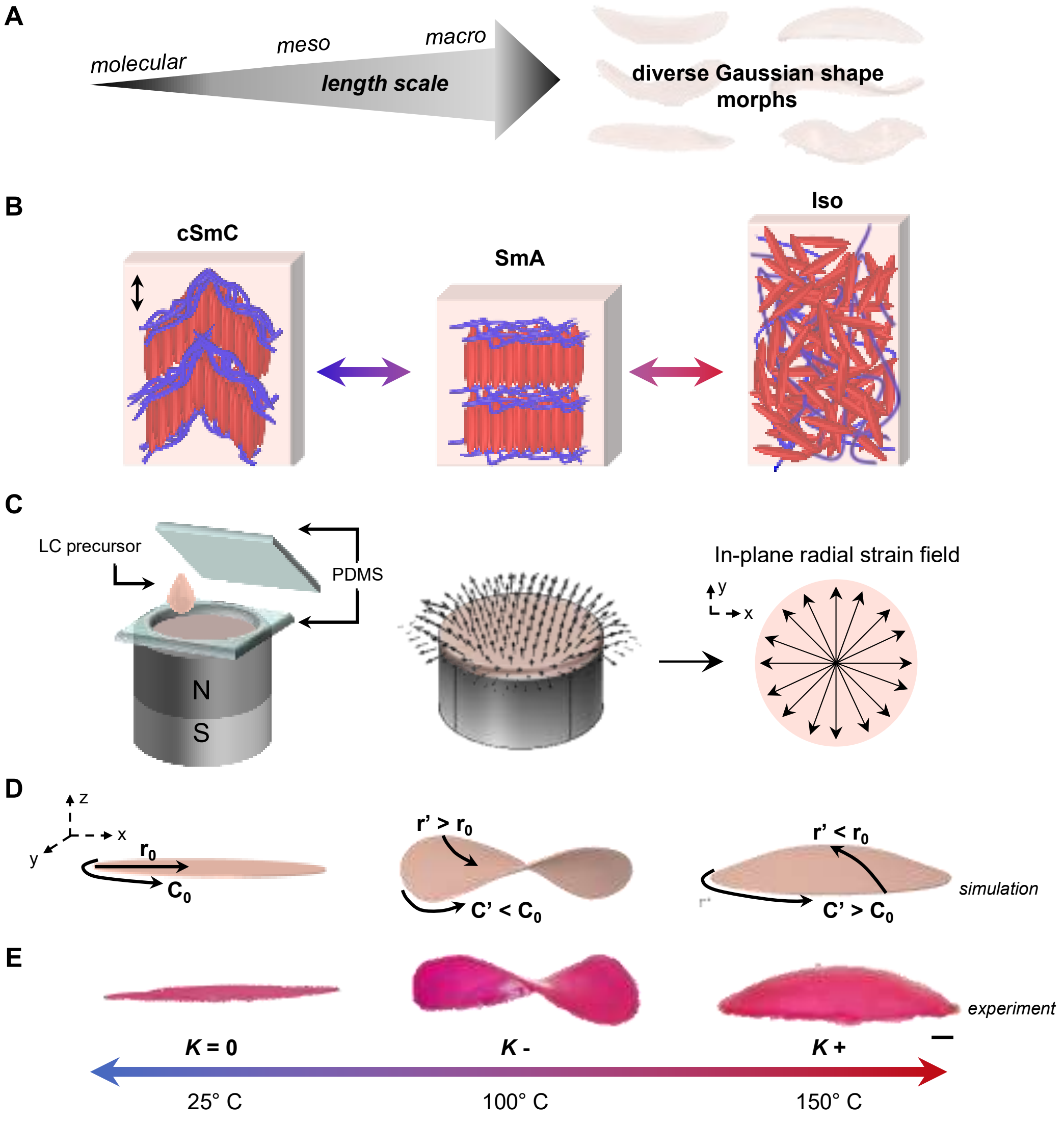

***Fig. 1) Ambidirectional LCE sheet displays opposite Gaussian curvature with monotonically increasing temperature.*** (**A**) Generating a diverse range of dynamic Gaussian shape morphs through hierarchical design from the molecular to the meso and macro scale. (**B**) Schematic of molecular packing of LC mesogens 4-(6-acryloxy-hex-1-yl-oxy)phenyl 4-(hexyloxy)benzoate in the chevron smectic C (cSmC), smectic A (SmA), and isotropic (Iso) phases. Pink rods represent LC mesogens, while blue lines represent polymer chains. The black arrow represents the axis of the director. (**C**) Fabrication scheme for preparing magnetically aligned radially patterned LCE sheets via soft lithography using polydimethylsiloxane (PDMS) molds. (**D**) Finite element simulations and (**E**) experimental results of ambidirectional LCE sheets displaying opposite Gaussian curvature upon continuous heating. $C_0$ and $r_0$ are the circumference and radius in the reference state, while *C'* and *r'* are the circumference and radius in the deformed state.

## 2.2 Curvature magnitude modulation through geometric tuning

Magnetic alignment patterns the director field throughout the sheet volume, while the multistep inversion of Gaussian curvature arises from the LCE's intrinsic molecular response rather than from multilayer construction or hybrid composite design. We therefore sought to isolate how molecular alignment and macroscale geometry jointly govern out-of-plane morphing in a single material. To make this relationship explicit, we focused on two fabrication parameters that directly tune curvature magnitude: sheet thickness, $t_{\text{sheet}}$, and the relative alignment radius, $R$, defined as

$$R = \frac{r_{sheet}}{r_{magnet}} \qquad eqn. 1$$

where $r_{sheet}$ and $r_{magnet}$ are the radii of the LCE sheet and magnet, respectively. At fixed $t_{\text{sheet}}$, curvature magnitude can be tuned through $R$. Conversely, when the cross-sectional area is prescribed, the thickness serves as an out-of-plane control parameter for tuning magnitude. Together, these two parameters provide distinct geometric control over shape change. For a fixed in-plane footprint, $t_{\text{sheet}}$ determines the sheet's bending resistance. For a fixed $t_{\text{sheet}}$, $R$ governs the fraction of the sheet experiencing tilted versus predominantly vertical director field (*Fig. S2*).

A sheet's resistance to bending scales with its thickness as $D \propto t_{sheet}^3$, where $D$ is the flexural rigidity[25–27]. Therefore, increasing $t_{sheet}$ increases the flexural rigidity, suppressing out-of-plane deformation and leading to $K \rightarrow 0$ for both phase transitions. We verified this relationship by simulating radially aligned sheets with increasing thickness and calculating the resulting Gaussian curvature during contraction and expansion, respectively (*Fig. 2A*). We next examined $r_{magnet}$ as a secondary control parameter. We simulated LCE sheets ($r_{sheet}$ = 2.5 mm, $t_{sheet}$ = 100 µm) aligned atop magnets with varying radii (*Fig. 2B*). Non-zero Gaussian curvature arose due to opposing stretch along the radial and circumferential direction. When $r_{magnet} \gg r_{sheet}$ or $R \ll 1$, the director field remained predominantly vertical toward the center of the magnet, resulting in nearly zero curvature, -0.00083 $mm^{-2}$ and 0.00078 $mm^{-2}$ during the cSmC-SmA and SmA-Iso phase transitions, respectively. As $r_{magnet}$ decreased, the director field developed larger in-plane alignment, increasing the strain anisotropy and thus, the magnitude of $K$. We note that for $r_{magnet}$ = 0.85 mm (R = 2.94) and 1.5 mm (R = 1.67), simulations of the cSmC-SmA transition resulted in a negative stiffness matrix and failed to converge due to the large strains. To improve numerical stability, we decreased the Poisson ratio to ν = 0.45. This modification reduced the predicted curvature magnitude. A Poisson ratio closer to incompressibility ($\nu = 0.5$), would produce larger curvature magnitudes.

After investigating the independent effects of the relative alignment radius and sheet thickness on the in-plane director field and *K*, we experimentally examined how these two control parameters interact to generate curved surfaces of varying magnitudes. Specifically, we varied R by modifying the $r_{sheet}$ while keeping $r_{magnet}$ fixed at 4 mm and $t_{sheet}$ fixed at 230 μm. After fabrication, we characterized the Gaussian curvature of sheets with $R = 0.4$, 1, 2, and 3 by performing micro-computed tomography (μCT) measurements for both the cSmC-SmA and SmA-Iso phase transitions (*Fig. 2C, Fig. S2*). The result of these measurements reveal a nonlinear relationship between *R* and *K* (*Fig. 2D*). When the $r_{sheet} < r_{magnet}$ ($R = 0.4$), the surface does not curve significantly during either the cSmC-SmA transition ($K = -0.0002\ mm^{-2}$) or the SmA-Iso transition ($K = 0.0001\ mm^{-2}$). When $r_{sheet} = r_{magnet}$ ($R = 1$), the magnitude of curvature during phase transition increases by two orders of magnitude ($K = -0.013\ mm^{-2}$ and $0.0067\ mm^{-2}$, respectively). However, further increasing R does not lead to a monotonic increase in Gaussian curvature. At $R = 2$, the magnitude of Gaussian curvature decreases for the cSmC-SmA transition ($K = -0.008\ mm^{-2}$) compared to $R = 1$ but increases for the SmA-Iso transition ($K = 0.012\ mm^{-2}$). At $R = 3$, curvature during both the cSmC-SmA and SmA-Iso transitions decreases compared to both $R = 1$ and $R = 2$ ($K = -0.003\ mm^{-2}$ and $0.0060\ mm^{-2}$, respectively). We attribute this non-monotonic behavior to two competing effects which arise when *R is varied*: 1) a change in the area with tilted alignment, and thus, in-plane strain, across the sheet and 2) a change in sheet aspect ratio which changes the sheet's global resistance to bending[27].

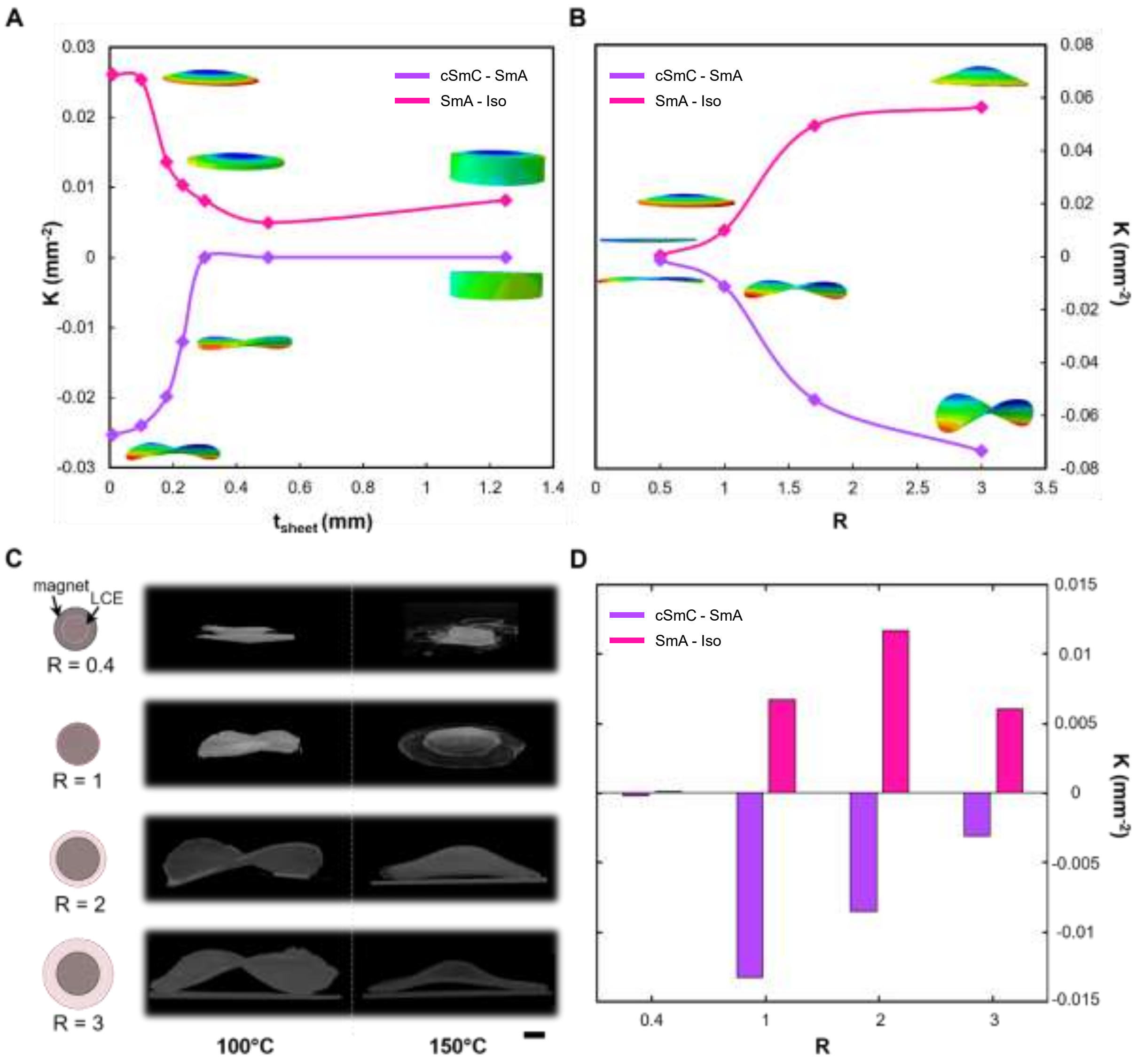

***Fig. 2) Mesoscale and geometric parameters for modulating Gaussian curvature magnitude.*** (**A**) Gaussian curvature (K) as a function of $t_{sheet}$ for an LCE sheet simulated with *R = 1*. *K* values were calculated by imaging the deformed shapes along their principal axes and evaluating each profile as a series of points. Fitting a circle to these points yields radii of curvature which are used to calculate *K*. (**B**) Gaussian curvature (K) as a function of relative alignment radius, *R*, for an LCE sheet simulated with $t_{sheet}$ = 0.1 mm. (**C**) micro-computed tomography (μCT) measurements of the deformed shape at the two LC phase transitions for sheets of varying values of R. (**D**) Gaussian curvature values at varying values of *R* after the cSmC-SmA and SmA-Iso phase transitions. Scale bar is 1 mm.

### 2.3 Encoding asymmetry in macroscopically symmetric sheets

By taking advantage of mesoscale asymmetry, we demonstrate a single geometrical form factor – a thin sheet – which exhibits a diverse range of oppositely curved deformed surfaces upon phase change. LCE sheets were polymerized atop a disk magnet with the sheet center offset from the center of the magnet. We define the center-to-center distance between sheet and magnet:

$$C = \frac{x_{sheet} - x_{magnet}}{r_{magnet}} \qquad eqn. 2$$

where $x_{sheet}$ and $x_{magnet}$ are the center positions of the sheet and magnet, respectively. When the magnet and sheet centers coincide ($C = 0$), the sheet morphs into a saddle and a dome as discussed previously (*Fig. 3A*). Offsetting the center of the sheet by $x_{sheet} - x_{magnet} = r_{magnet}$ ($C = 1$) induces the same shape morphing behavior as $C = 0$, except that the inflection points, denoted by blue asterisks, are no longer positioned at $x_{sheet}$, denoted by red asterisks (*Fig. 3B*). Further, during the SmA-Iso transition, the sheet develops a second inflection point along its centerline. We associate this secondary point with the horizontal alignment in the region approaching the edge farthest from the magnet. We posit that, because the sheet is softest at 150°C, this alignment is sufficient to induce additional curvature at this temperature (*Fig. S4*).

For a sheet offset by $2 \cdot r_{magnet}$ ($C = 2$), the cSmC-SmA transition produces nearly zero out-of-plane bending because the director field is aligned almost completely in the horizontal direction (along the x-axis) (*Fig. 3C*). Interestingly, further heating to the isotropic phase leads to curling of the sheet into a surface of zero Gaussian curvature. Like the isotropic case for $C = 1$, we attribute this large out-of-plane deformation to the combination of inhomogeneity in the director field and softening of the sheet. These results demonstrate that offsetting the sheet within the magnetic field of a single magnet produces asymmetry along the offset axis. A second strategy for producing director fields which break radial symmetry is to polymerize sheets in a magnetic field produced by two adjacent magnets. When LCE sheets are centered atop two magnets with their poles oriented in the same direction (i.e., both north poles facing up), the like-pole configuration generates a left-right symmetric director field, and the sheet exhibits saddle-dome shape morphing behavior like the mesoscopically axisymmetric sheets (*Fig. 3D*). However, when the two magnets are aligned with their poles oriented in different directions (i.e., one north and one south pole facing up), the field is primarily horizontal between the magnets. Initially, the sheet exhibits minimal out-of-plane deformation before morphing into a curved surface with two regions of positive Gaussian curvature connected by a horizontally deforming central region (*Fig. 3E*). Thus, spatially asymmetric magnetic fields can transform a single disk into a versatile platform for programming curvature, bending symmetry, and deformation magnitude.

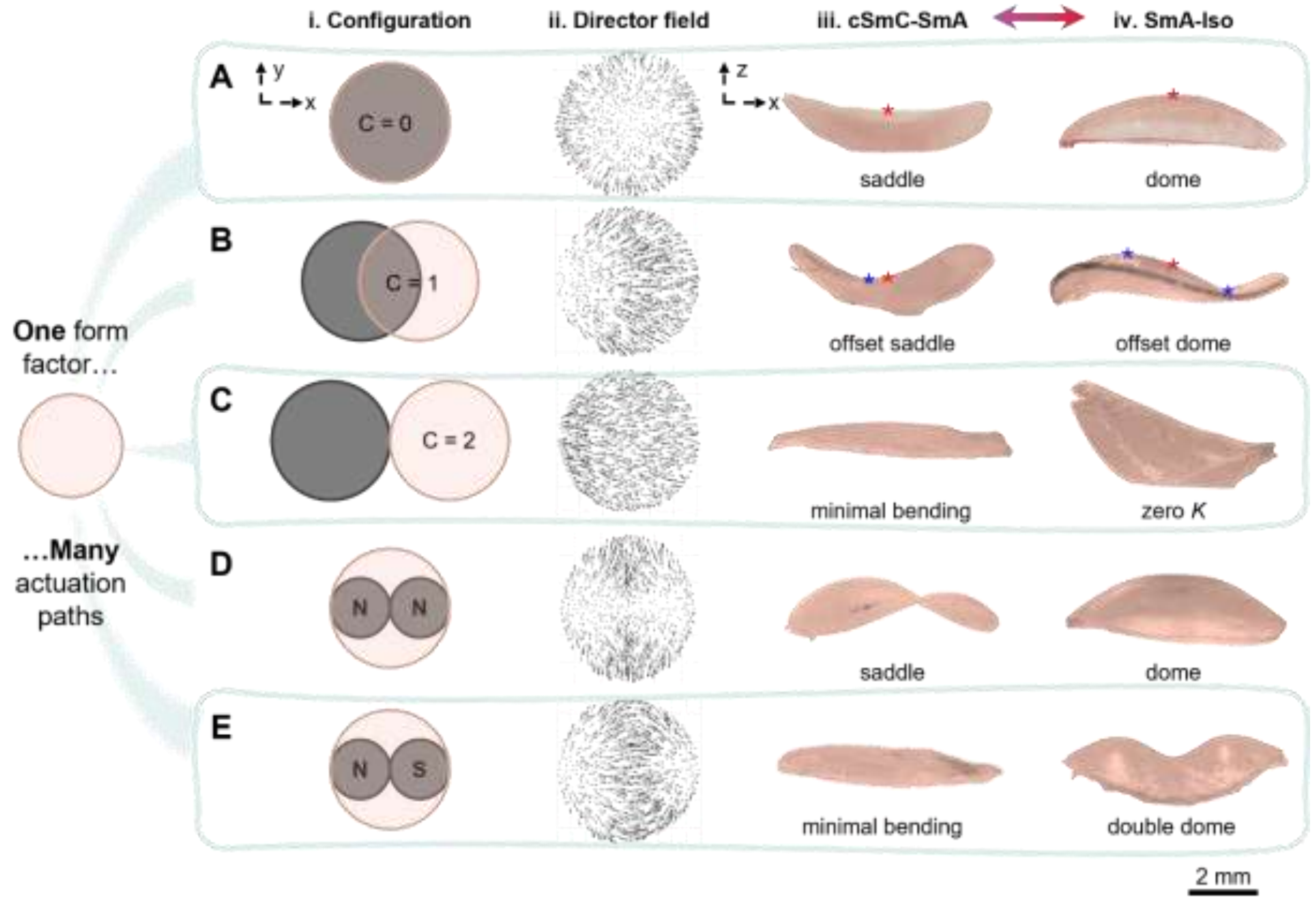

***Fig. 3) Ambidirectional Gaussian curvature in radially patterned sheets of varied mesoscale symmetry.*** (**A-C**) fabrication configuration (i.), director field (ii.), and resulting deformation upon heating to the SmA (iii.) and Iso (iv.) phases for sheets aligned with C = 0 (**A**), C = 1 (**B**), and C = 2 (**C**). (**D & E**) Deformation of sheets aligned atop two magnets, with either both north poles facing up (**D**) or one north pole facing up and one south pole facing up (**E**).

### 2.4 Dynamic optical response in ambidirectional soft photonic actuators

We next exploited this programmability to couple mechanical compliance with optical response, a phenomenon observed in many soft-bodied marine organisms, including octopuses, squid, and cuttlefish, which actively use coloration as a means of camouflage and interspecies communication[28,29]. In cephalopods such as the *Doryteuthis pealeii,* this capability arises from the combined action of iridophores – colorless cells containing thin plates which reflect light via thin-film interference and chromatophores – pigment-filled sacs surrounded by radially aligned muscle fibers[30]. Upon neural stimulation, radial muscles within the skin deform these cells to modulate their angle of illumination and other structural properties, resulting in dynamic coloration. Inspired by this color-changing mechanism, we coupled our ambidirectional out-of-plane deformations to dynamic optical properties via structural coloration (*Fig. 4A*). Structurally colored LCE sheets were fabricated by polymerizing precursor mixture in PDMS molds where one side was covered with a PDMS diffraction grating. Scanning electron microscopy (SEM) revealed a continuous square-wave surface grating consisting of 1 μm wide grooves and ridges, corresponding to a 2 μm pitch (500 grooves $mm^{-1}$) (*Fig. S5*).

We investigated the color-changing response as a function of phase change by observing the sheet at a fixed viewing angle during heating. Sheets were placed onto the surface of a hot plate and imaged using a digital single-lens reflex camera attached to a tripod. To ensure the sheet was free to deform without sticking to the surface of the hot plate, the featureless side of the sheet was coated with a thin layer of graphene oxide. This coating consequently turned the sample black, providing good contrast for the reflected light during imaging. In the first setup (*Fig. 4B*), we imaged the entire surface of the sheet during the cSmC-SmA-Iso phase transition. At this viewing angle, the sheet was predominantly black. Upon heating, the position-dependent diffraction response shifted as the sheet bent out-of-plane, thereby varying the grating orientation and causing localized regions to become reflective[31]. After morphing into a saddle shape, the number and position of these reflective regions changed, producing localized red and green iridescent patterns. Further heating led to the reflection spots shifting until the sheet fully morphed into a dome, at which point there was a new pattern at a different location across the surface. We then examined the same sheet from a second viewing angle at higher magnification to better resolve local changes in optical response. At this viewing angle, the selected region was initially bright blue. Inducing the saddle and dome shape morphs causes this viewed region to shift across the color spectrum before losing all reflectance in the dome state (*Fig. 4C*). For this viewing angle (*Fig. 4C*), we quantified changes in the R, G, B channels during phase transitions and captured a highly dynamic, color-changing process (*Fig. S6*). These results demonstrate a strategy to use multistep shape-mediated angular tuning to produce dynamic, spatially varied iridescence. This strategy allows for encoding angle-dependent optical behavior as a means of storing a wealth of optical and structural information across the surface of a dynamically curved sheet. The color-changing behavior observed in these experiments can be further tuned through incorporation of the geometric designs explored within this work and by modulating the heating rate. Ultimately, these surface-textured shape morphs may offer promising opportunities in applications such as information-encrypted surfaces and color-changing motile robots.

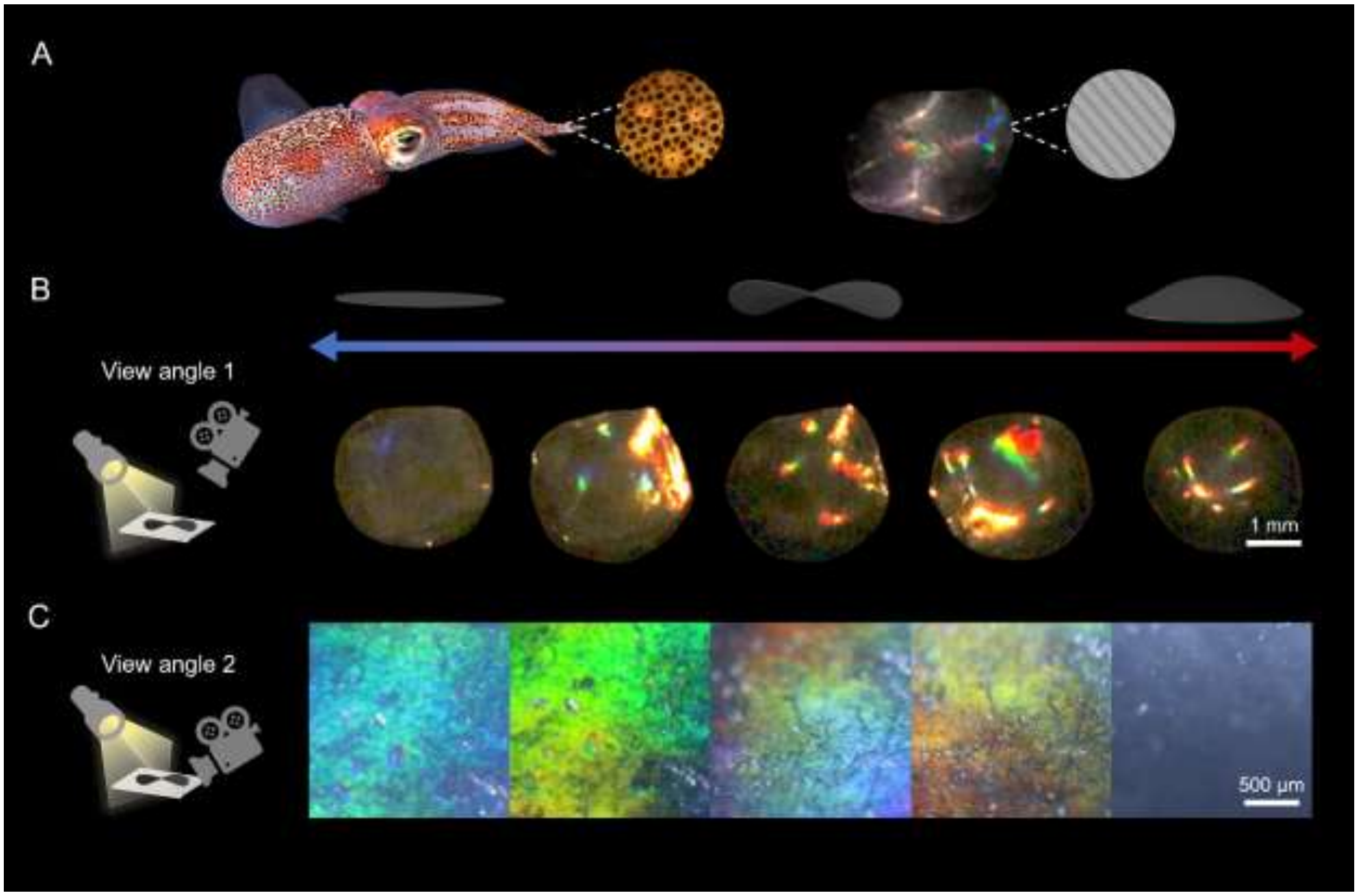

**Fig. 4) Bioinspired soft photonic actuators.** (**A**) (left) An adult Hawaiian bobtail squid (*Euprymna scolopes)* exhibiting surface patterns on their skin due to structural and pigmented coloration. Hawaiian bobtail squid image by Margaret McFall-Ngai, via Wikimedia Commons, licensed under CC BY-SA 4.0. Chromatophore inset image by Minette, via Wikimedia Commons, licensed under CC BY 2.0. Images cropped and arranged for this figure. (right) An LCE sheet displaying patterns along its surface due to a diffraction grating (2 µm pitch (500 grooves $mm^{-1}$). (**B**) A radially aligned LCE sheet with R = 1 exhibiting ambidirectional surface curvature coupled to dynamic optical changes upon heating from 25°C to 150°C at fixed viewing angle 1. (**C**) Magnified view of dynamic optical changes at fixed viewing angle 2, showing color change of an initially uniform spot.

## 3. Conclusion

This work establishes a hierarchical design framework for programming multistate Gaussian shape transformations in ambidirectional LCE sheets. By coupling phase transition-enabled strain inversion with spatially encoded director fields, we show how a single-sheet architecture can be designed to access distinct thermally driven shape morphs under monotonically increasing temperature. We further identify how geometric parameters, including sheet thickness, relative alignment radius, and offset within the magnetic field, govern the magnitude, symmetry, and spatial distribution of curvature, thereby providing practical design guidance for tuning the resulting morphologies. Coupling these deformation pathways with surface diffraction further enables dynamic optical behavior, linking multistate shape reconfiguration to tunable structural coloration. Together, these findings demonstrate the promise of coordinating molecular response, alignment pattern, and macroscale geometry to encode mechanical and optical function within a single-material system. Such capabilities may open new opportunities for responsive surfaces and devices in fields such as soft robotics, adaptive optics, camouflage, and information-rich photonic actuation.

**Supporting Information**

The online version contains supplementary material available as ancillary files alongside this article on arXiv.

**Acknowledgments**

This work was supported by the National Science Foundation (NSF) through the Harvard University Materials Research Science and Engineering Center (MRSEC) under the award DMR-2011754. Micro-computed tomography and atomic force microscopy were performed at the Center for Nanoscale Systems (CNS) at Harvard, a member of the National Nanotechnology Coordinated Infrastructure Network (NNCI), which is supported by the NSF under NSF ECCS Award No. 1541959.

**Author Contributions**

A.M.W. and J.A. conceived the idea and designed the study. A.M.W., G.K.P., R.K., P.W., F.S., J.M., and S.L. performed research. A.M.W. and G.K.P. analyzed the data. A.M.W. and J.A. wrote and revised the paper.

**Data Availability Statement**

The data that support the findings of this study are available from the corresponding author upon reasonable request.

**Conflict of Interest**

The authors declare no conflict of interest.

Supplementary Information for

# Hierarchical Design of Ambidirectional LCE Sheets for Three-State Gaussian Shape Morphs

Atalaya Milan Wilborn[1,2]; Gurminder K. Paink[1]; Raphael Kay[1]; Patryk Wąsik[3]; Friedrich Stricker[1]; Jacopo Movilli[1]; Shucong Li[4,5], Joanna Aizenberg[1,5]*

1. John A. Paulson School of Engineering and Applied Sciences, Harvard University, Cambridge, MA 02138, USA
2. Complex Materials, Department of Materials, ETH Zürich, Zürich, 8093 Switzerland
3. National Synchrotron Light Source II, Brookhaven National Laboratory, Upton, New York 11973, USA
4. School of Materials Science and Engineering, Georgia Institute of Technology, Atlanta, GA, USA
5. Department of Chemistry and Chemical Biology, Harvard University, Cambridge, MA, USA.

*Corresponding author. Email: jaiz@seas.harvard.edu.

This Supplementary Information document includes:

Materials and Methods

Figures. S1 to S6

## Materials and methods

Materials

The end-on LC monomer 4-(6-acryloxy-hex-1-yl-oxy)phenyl 4-(hexyloxy)benzoate was purchased from Synthon Chemicals. Photoinitiator Darocur 1173 (2-hydroxy-2-methylpropiophenone), anhydrous dichloromethane (DCM), and silicone oil (10,000 cSt) were purchased from Merck/Sigma Aldrich and used as received. The crosslinker 1,6-hexanediol diacrylate (HDDA, from Sigma Aldrich) was used after removal of inhibitor. Methacryloxyethyl thiocarbamoyl rhodamine B dye was purchased from Polysciences, Inc. Sylgard 184 silicone kit was purchased from Ellsworth Adhesive Systems. High temperature neodymium (NdFeB) magnets (N42SH grade) with different shapes and dimensions were purchased from K&J Magnetics, Inc.

LCE sheet fabrication

Negative PDMS molds for casting LCE sheets were fabricated using a previously reported method[1]. Briefly, macroscopic geometries were generated by laser cutting and assembling Artus plastic shim stock with desired size and thickness in a petri dish, followed by PDMS casting to make molds with negative features. The obtained PDMS molds determined the geometry of the LCE sheets.

For LCE fabrication, the negative PDMS mold was placed on top of a setup of magnets heated on a hot plate set to 80°C, with a surface magnetic field of ~0.5 T. About 25 mg of LC precursor mixture was sandwiched between the negative PDMS mold and a flat PDMS sheet backed by a glass coverslip to ensure rigidity. The mixture was then cooled from 80°C to 65°C at a rate of 1°C/min. To ensure that the temperature was uniform throughout the mixture, the sample was held at 65°C for 1 hour before being cooled further to 40°C at 1°C/min. UV-initiated polymerization was carried out for 1 hour using a Dymax Model 2000 Flood UV Curing System with a light intensity of approximately 18 mW/cm$^2$, under nitrogen gas protection. After polymerization, the LCE film was carefully peeled out of the PDMS mold.

LCE diffraction grating fabrication

To fabricate LCE sheets with diffraction gratings on the surface, we modified our soft-lithography technique as follows: LC precursor mixture was sandwiched between a negative PDMS mold and a PDMS diffraction grating. To ensure complete filling of the melted LC precursor into the grating features, we placed the sample under vacuum at 80°C for 30 minutes. Afterward, we placed the sample atop a disk magnet, centered on the center of the magnet ($R = 1$, $t_{sheet} = 0.23$ µm), aligned and polymerized according to our previously established method. Upon removal of the sample from the PDMS sandwich, the diffraction grating features were successfully imprinted into the surface of the LCE sheet.

Scanning electron microscopy

Scanning electron microscopy (Zeiss Gemini 360 FE-SEM) (SEM) was used to image the surface of the LCE diffraction grating sheets. A small piece of LCE sheet was mounted onto a SEM stub (Ted Pella) using double sided conductive carbon tape. A 10 nm Pt/Pd conductive coating was subsequently sputtered onto the mounted samples using an EMS150T S sputter coater.

Atomic force microscopy

Atomic force microscopy (Bruker JPK Nanowizard) was used to investigate the mechanical properties of the LCE sheet. Scans were performed on cross-sections (parallel alignment to compressive load) with a 10 um x 10 um sampling area and 32x32 pixel resolution in QI™ mode using all-in-one-Al probes (Budget Sensors) with a spring constant of 0.2-2.7 N $m^{-1}$ and resonant frequency of 15-80 kHz. Samples were allowed to equilibrate for 15 minutes at temperature before measurements. Measurements were performed at 25ºC, 100ºC and 180ºC. An n = 3 measurements were performed at each temperature. The resulting scans were analyzed using the JPK analysis software to measure the average modulus values in each scan.

X-ray scattering

Wide angle x-ray scattering experiments to characterize the mesogen alignment conducted at the 12-ID beamline of the National Synchrotron Light Source II. The incident x-ray beam was focused to a spot size of 25 µm (horizontal) × 2.5 µm (vertical) with energy of 16.10 keV at the sample position. The samples were mounted on to a Linkam HFSX350-GI heating/cooling stage inside of a vacuum chamber. Measurements were taken perpendicular to the mesogenic director in transmission mode and recorded using a PILATUS3 900 K-W detector with an exposure time of 0.5 seconds per frame. To exclude any surface effects, measurements were taken at a spot far enough from the sample surface (>100 µm away).

Finite element simulations

To model the spontaneous deformation of LCEs upon phase change, we employed a UMAT with the following constitutive equation[1,2]:

$$W = \frac{1}{2} C_{ijkl}\varepsilon_{ij}\varepsilon_{kl} - \beta(S^t - S^0)\frac{1}{2}\left(3n_i n_j - d_{ij}\right)\delta_{ij} \qquad (eqn. 3)$$

where $W$ is the total strain energy of the elastomer, $C_{ijkl}$ is the stiffness tensor for an isotropic rubber, built from the Young's modulus and Poisson's ratio of the material, $\hat{n}$ is the director vector with components $n_i$ and $n_j$, $\beta$ is a coupling parameter used to adjust LC ordering against classical elasticity, and $S^t$ and $S^0$ are the scalar order parameter in the current and reference states, respectively. During a static general step analysis in Abaqus, we incrementally sweep the value of $S$ from 1 to 0, with nodal forces/displacements calculated by minimizing the free energy equation at each increment.

To phenomenologically capture the mechanical behavior of end-on LCEs undergoing two phase transitions, we modified the sign of $\beta$ to be positive for simulating the first phase transition (cSmC–SmA) and negative for the second phase transition (SmA–Iso). To obtain the minimum free energy state, a positive $\beta$ penalizes positive strain along $\hat{n}$, effectively simulating uniaxial contraction of the element. A negative $\beta$ therefore encourages positive strain along $\hat{n}$, leading to uniaxial expansion of the element.

End-on LCE sheets with inhomogeneous director orientations were simulated by first defining the desired geometry and discretizing into a mesh of elements. Subsequently, we generated a list of global centroidal positions for each element calculated from the nodal coordinates and connectivity of the mesh. These centroids served as reference points for assigning the correct director orientation based on their proximity to magnetic flux vectors. To encode the proper director field for each sheet geometry and magnetic field configuration, we compared the positional data of elements in the mesh with corresponding normalized magnetic flux vectors across the surface of the magnet. Magnetic flux vectors were obtained by simulating configurations of 0.2 T axially aligned permanent disk magnets in COMSOL Multiphysics. A list of 90,000 equally spaced flux vectors along the top surface of each magnet configuration was then exported and used to assign the local mesogenic alignment.

Gaussian curvature determination

To determine the 3D configuration of LCE sheets in their deformed states, we performed micro-computed tomography (μCT) measurements using a Nikon Metrology HMX ST 225 μCT scanner. To perform these measurements at room temperature, we developed the following method for preserving each deformation state: the sheets were heated in a silicone oil bath at 100°C and 150°C corresponding to the SmA and Iso phase, respectively and subsequently quenched by rapid submersion in a bath of dry ice and acetone.

After measurement and 3D reconstruction, the planes containing the principal axes are determined by finding the axis which contains the point of maximum curvature in the XY-plane and enforcing orthogonality to reach the axis containing the point of minimum curvature (*Fig. S2A*). The XZ and YZ-planes at these locations are then imaged, with points along each curve plotted as line profiles. After imaging the sheets along their X-Z and Y-Z planes, defined as the planes containing the axis of principal (minimum and maximum) curvature, (*Fig. S2A*), points along the bottom edge of the sheets (presumed to be the central axis of the disk) were plotted in MATLAB to obtain line profiles (*Fig. S2B, left*). Subsequently, we calculated the principal radius of curvature ($k$) along each direction with the following equation:

$$k = \frac{1}{R} \qquad eqn. (S1)$$

where $R$ is the radius of a circle that fits the experimentally observed curved surface through least-squares fitting in MATLAB[3] (*Fig. S2B, right*).

**Supplementary Figures**

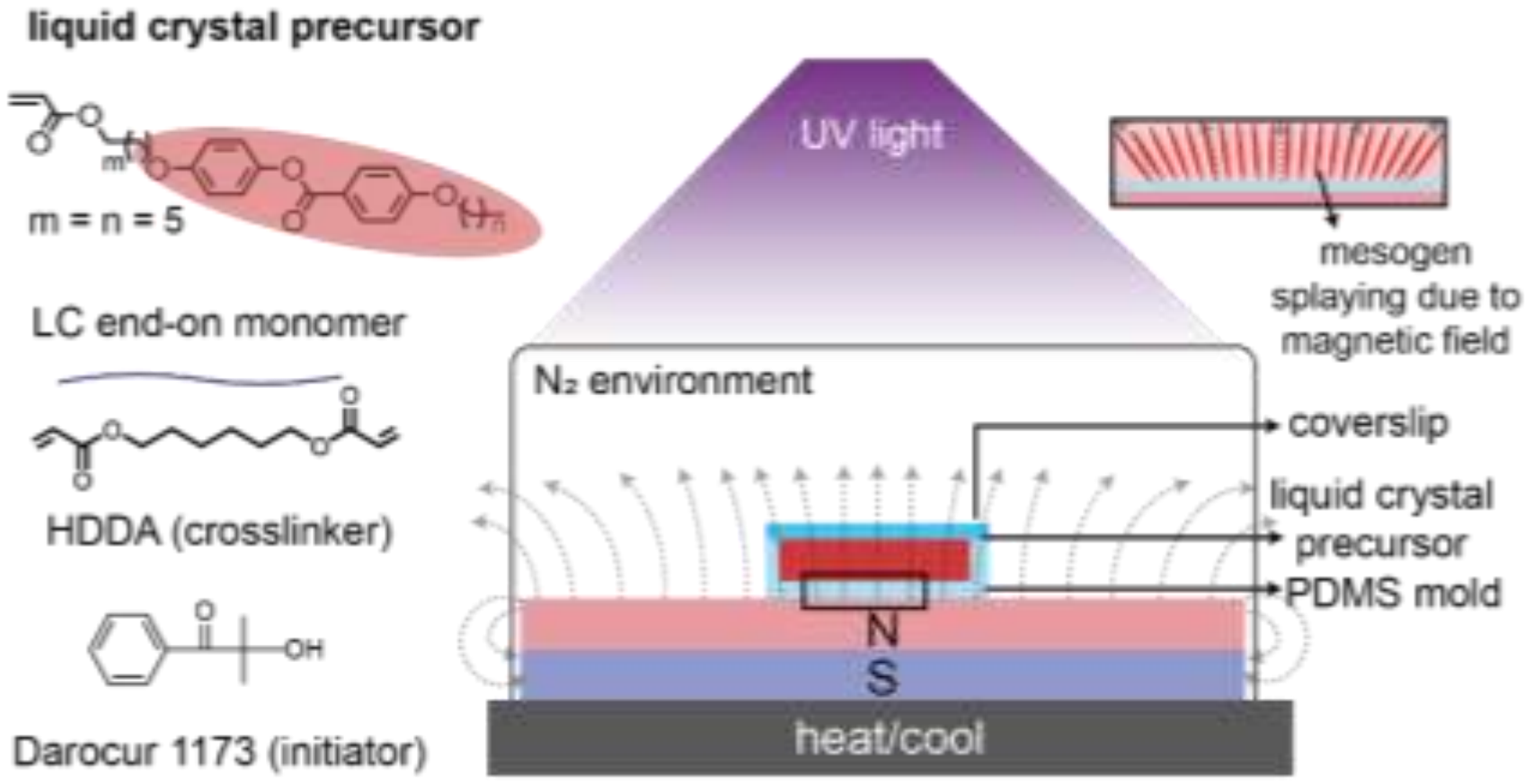

***Fig. S1) Schematic of LCE sheet fabrication.*** Precursor mixture comprised of end-on monomer, crosslinker, and photo initiator is cast into a PDMS mold atop a disk magnet and heated to 80°C. The mixture is cooled to 40°C before UV irradiation to initiate polymerization.

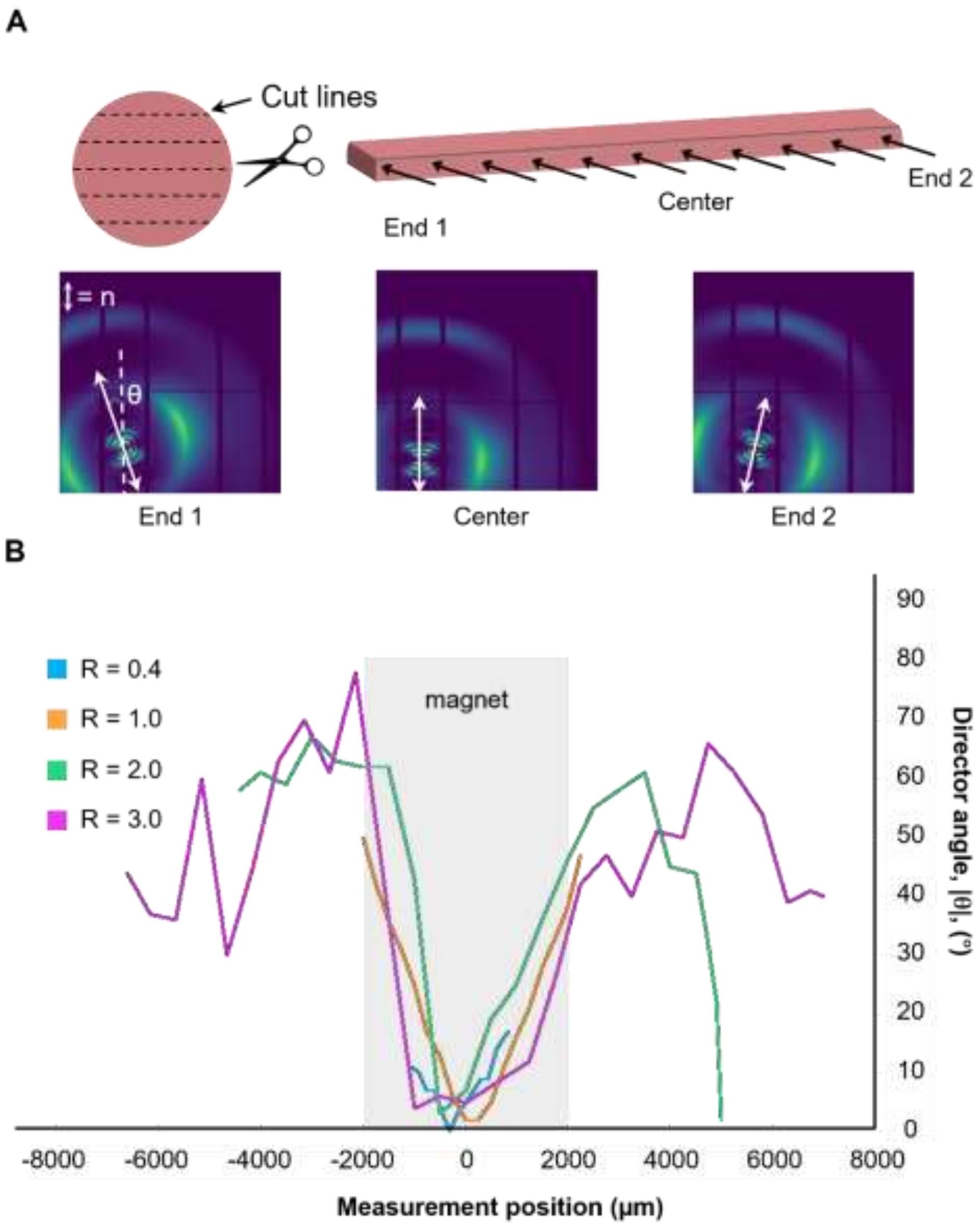

***Fig. S2) Director profiles across the length of end-on LCE sheets polymerized under varying values of R.*** (**A**) LCE samples were cut with a razor blade to obtain a 1 mm wide thick slices of the central region of each disk. Wide-angle X-ray scattering (WAXS) measurements were taken every 50 μm along the length of the slices. Black arrows denote the beam positioning for measurements (top). Representative WAXS measurements (bottom) reveal the axis of the director at the two ends and center of a slice, where *n* denotes the director orientation and θ represents the director angle relative to vertical. (**B**) Director orientations along the length of LCE slices for disks with varied values of *R*. Increasing R leads to an increase in ratio of tilted alignment to vertical alignment, as theta increases with increasing distance from the center.

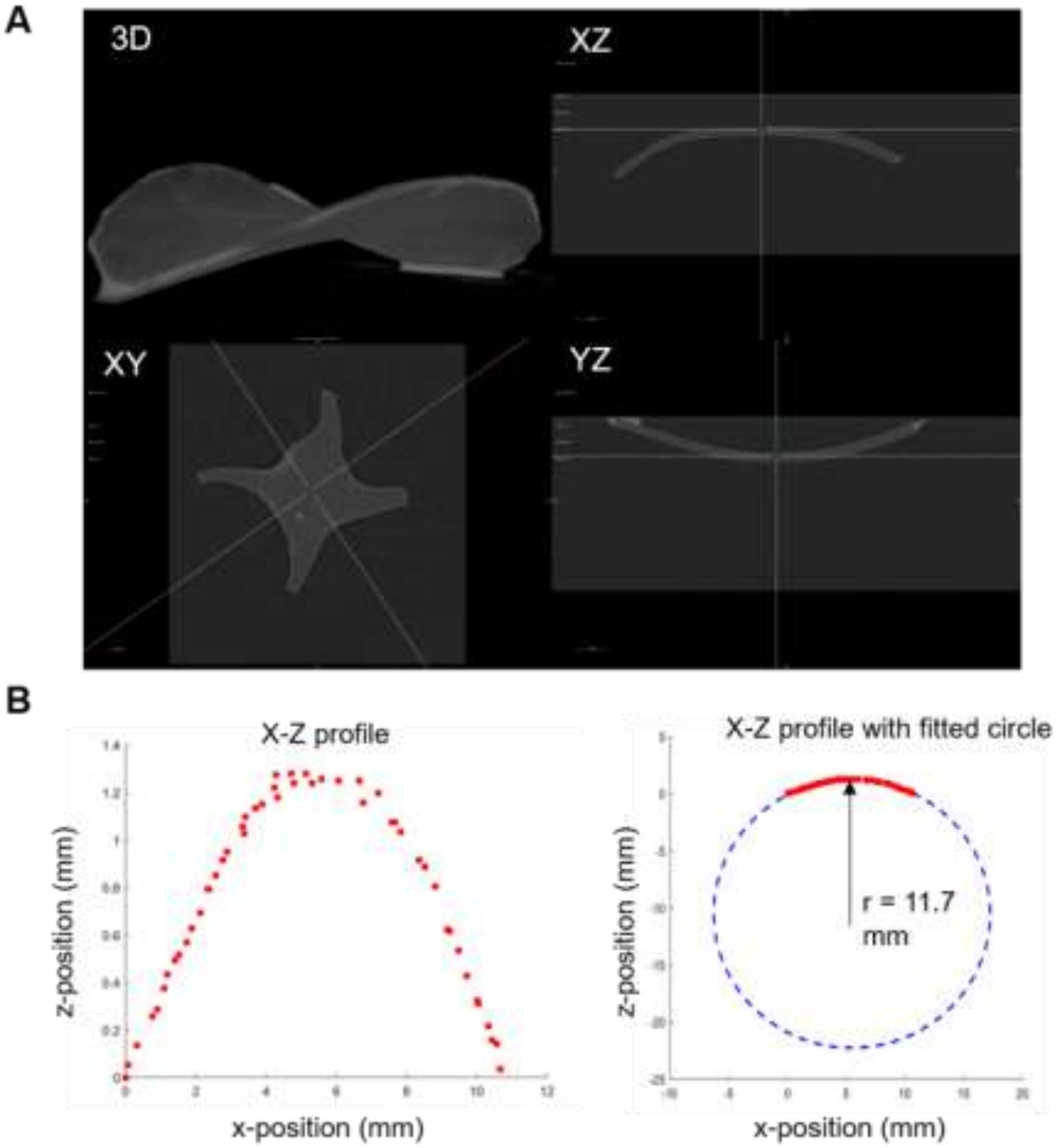

***Fig. S3) Representative data analysis of Gaussian curvature from µCT measurements.*** (**A**) After 3D reconstruction of the deformed shape, cross-sections are obtained in all three planes. The planes containing the principal axes are determined by finding the axis which contains the point of maximum curvature in the XY-plane and enforcing orthogonality to reach the axis containing the point of minimum curvature. (**B**) Line profiles for each curve in the XZ and YZ-plane are calculated and plotted (left). A circle is fitted to these values with a radius corresponding to the radius of curvature for this principal curve (right).

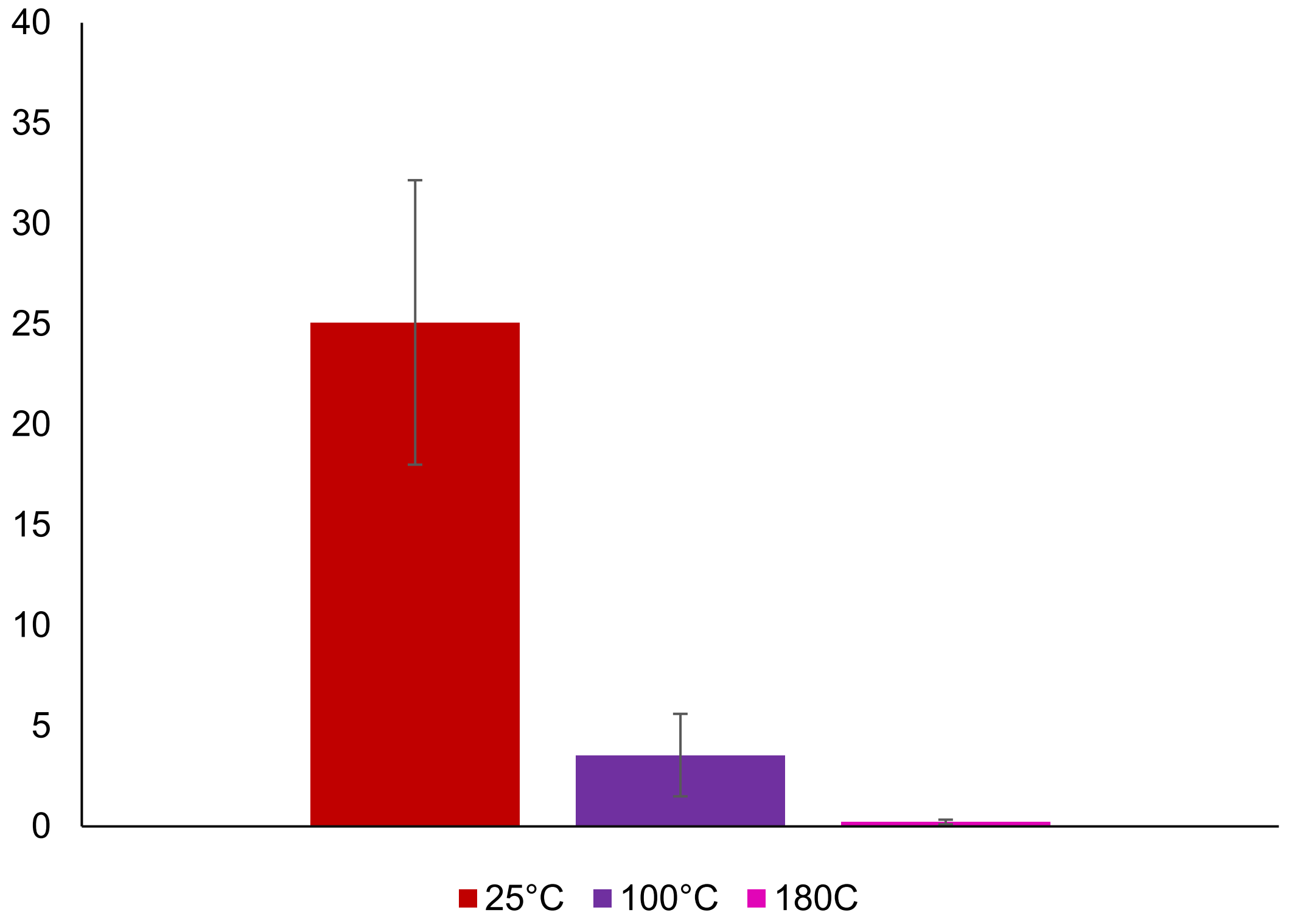

***Fig. S4) LCE Young's modulus as a function of temperature.*** Modulus values were obtained via atomic force microscopy (AFM). The sample was aligned parallel to the direction of compressive loading

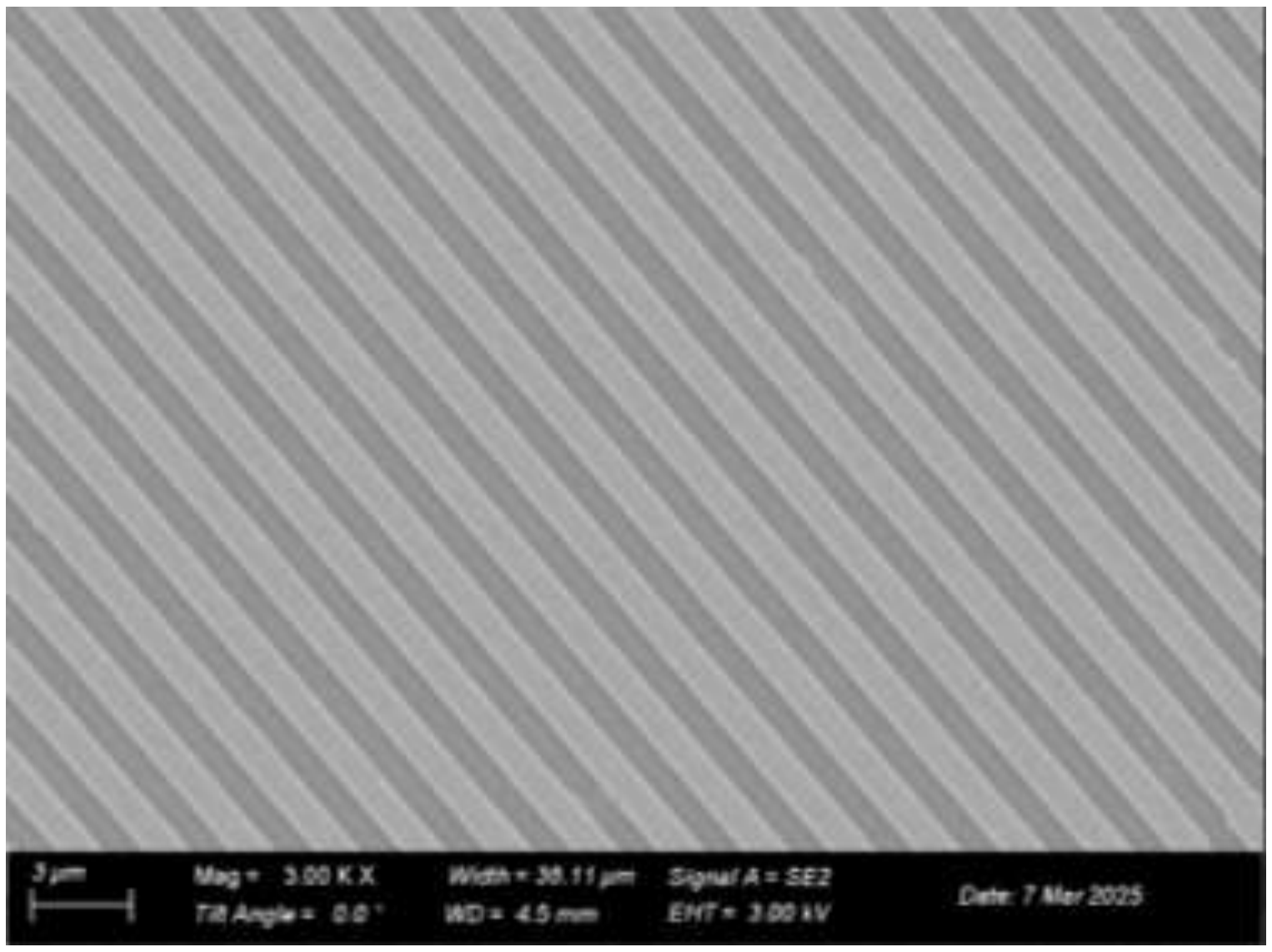

***Fig. S5) LCE diffraction grating characterization.*** SEM image of the surface of the LCE diffraction grating comprised of 1μm wide x 1μm pitch x 0.33 μm deep line features.

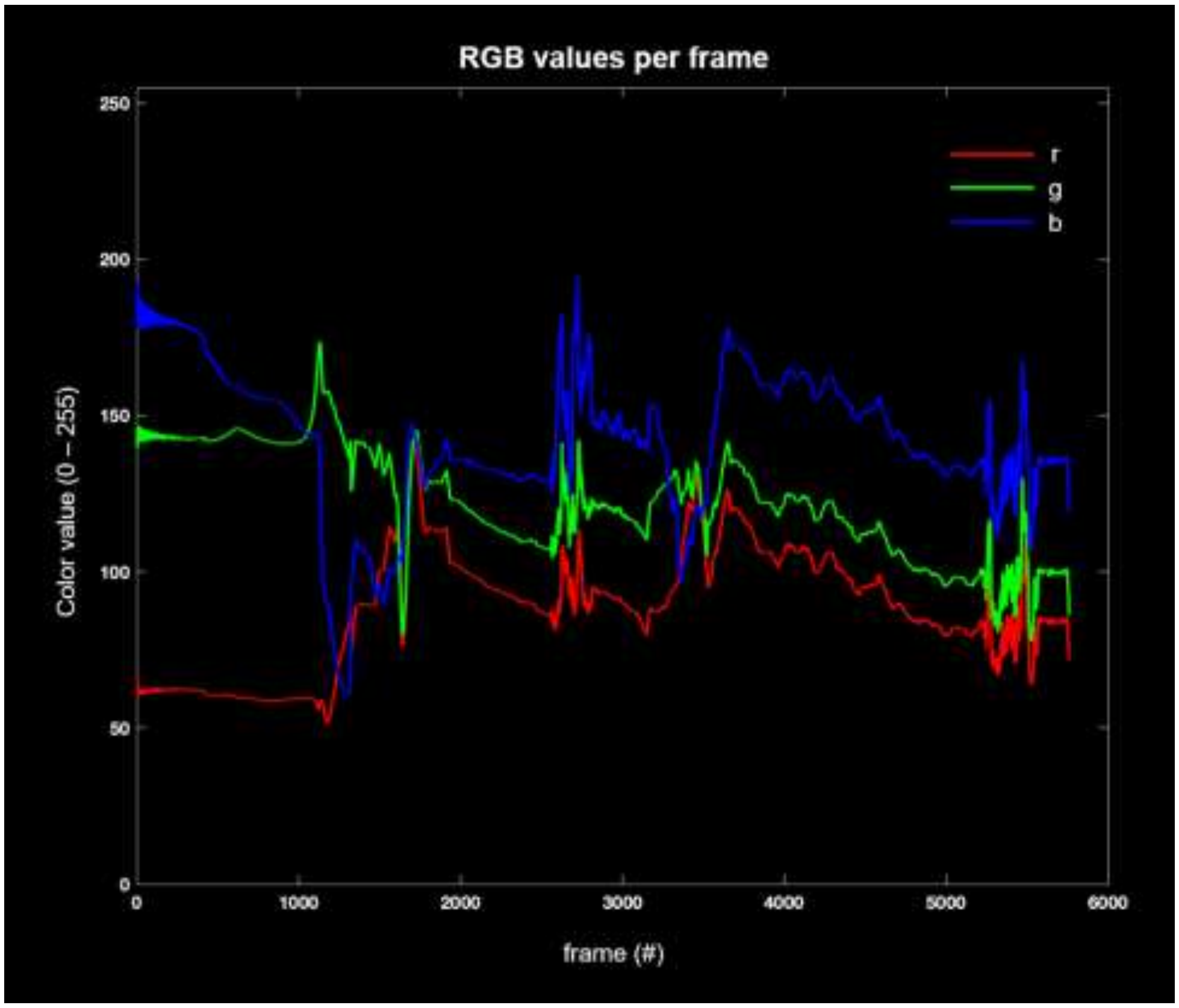

***Fig. S6) Dynamic color at viewing angle 2 upon heating from 25°C to 150°C.*** Red-green-blue values extracted from each frame of continuous recording for the LCE diffraction grating sheet at view angle 2.